\documentclass[12pt,preprint]{aastex}
\usepackage{natbib}
\usepackage{amssymb, amsmath, epsfig, epsf}  

\newcommand{\lum}{{\rm erg~s\ensuremath{^{-1}}}}

\newcommand{\ledd}{\ensuremath{L\mathrm{_{Edd}}}}
\newcommand{\lratio}{\ensuremath{L/\ledd}}

\newcommand{\kms}{\ensuremath{\mathrm{km~s^{-1}}}}

\newcommand{\rs}{\ensuremath{r_{\rm \scriptscriptstyle S}}}
\newcommand{\pnull}{ \ensuremath{P_{\mathrm{null}}} }

\newcommand{\ha}{{\rm H\ensuremath{\alpha}}}
\newcommand{\hb}{{\rm H\ensuremath{\beta}}}

\newcommand{\nii}{[N\,{\footnotesize II}]}

\newcommand{\oiii}{{\rm [O\,{\footnotesize III}]}}

\newcommand{\feii}{Fe\,{\footnotesize II}}
\newcommand{\crii}{Cr\,{\footnotesize II}}
\newcommand{\niii}{Ni\,{\footnotesize II}}
\newcommand{\tiii}{Ti\,{\footnotesize II}}

\newcommand{\mgii}{Mg\,{\footnotesize II}}

\def\lax{{$\mathrel{\hbox{\rlap{\hbox{\lower4pt\hbox{$\sim$}}}\hbox{$<$}}}$}}
\def\gax{{$\mathrel{\hbox{\rlap{\hbox{\lower4pt\hbox{$\sim$}}}\hbox{$>$}}}$}}

\slugcomment{{\it The Astrophysical Journal Letters}, 721, L143-L147 (2010) }
\shorttitle{Narrow optical \feii\ emission lines}
\shortauthors{Dong et al.}

\begin{document}

\title{The Prevalence of Narrow Optical \feii\ Emission Lines
in Type 1 Active Galactic Nuclei}

\author{
Xiao-Bo~Dong\altaffilmark{1,2}, Luis~C.~Ho\altaffilmark{2},
Jian-Guo~Wang\altaffilmark{3,1,5},
Ting-Gui~Wang\altaffilmark{1}, Huiyuan~Wang\altaffilmark{1}, \\
Xiaohui~Fan\altaffilmark{4}, and Hongyan~Zhou\altaffilmark{1}}
\email{xbdong@ustc.edu.cn}

\altaffiltext{1}{Key Laboratory for Research in Galaxies and Cosmology,
The University of Sciences and Technology of China, Chinese Academy of Sciences,
Hefei, Anhui 230026, China}
\altaffiltext{2}{The Observatories of the Carnegie Institution for
Science, 813 Santa Barbara Street, Pasadena, CA 91101, USA}
\altaffiltext{3}{National Astronomical Observatories/Yunnan Observatory and
Key Laboratory of the Structure and Evolution of Celestial Objects,
Chinese Academy of Sciences, Kunming 650011, China}
\altaffiltext{4}{Steward Observatory, The University of Arizona, Tucson, AZ 85721, USA}
\altaffiltext{5}{Graduate School of the Chinese Academy of Sciences,
19A Yuquan Road, P.O.Box 3908, Beijing 100039, China}

\begin{abstract}
From detailed spectral analysis of a large sample of low-redshift active
galactic nuclei (AGNs) selected from the Sloan Digital Sky Survey, we
demonstrate---statistically for the first time---that narrow optical \feii\
emission lines, both permitted and forbidden, are prevalent in type~1 AGNs.
Remarkably, these optical lines are completely absent in type~2 AGNs, across
a wide luminosity range, from Seyfert 2 galaxies to type~2 quasars.
We suggest that the narrow \feii-emitting gas is confined to a disk-like
geometry in the innermost regions of the narrow-line region on physical
scales smaller than the obscuring torus.
\end{abstract}


\setcounter{footnote}{0}
\setcounter{section}{0}

\section{Introduction}

Broad \feii\ multiplet emission are prominent features in the optical spectra
of most type~1 active galactic nuclei (AGNs), but narrow optical \feii\
lines, either permitted or forbidden, are rarely seen.  To our knowledge,
narrow optical \feii\ emission has been reported in only three sources to
date: I~Zw~1 (V\'eron-Cetty et al. 2004), Mrk~110 (V\'eron-Cetty et al.
2007), and SDSS\,J1028+4500 (Wang et al. 2008). The apparent weakness of
\feii\ emission in the narrow-line region (NLR) might be attributed to
iron, a refractory element, condensing onto dust grains, which sublimate in
the broad-line region (Laor \& Draine 1993).
This is especially true for low-ionization lines such as \feii\ that arise
from partially neutral portions of the NLR clouds (Ferguson et al. 1997).
On the other hand, narrow forbidden \feii\ emission lines in the near-infrared
(NIR) are observed commonly in both type~1 and type~2 AGNs (e.g., Simpson et
al.  1996; Mouri et al. 2000; Rodr{\'{\i}}guez-Ardila et al. 2004, 2005).
According to these studies, NIR [\feii] emission most likely originates
from AGN photoionization, although additional contributions from shock
heating cannot be excluded.  This raises the following questions: are narrow
optical \feii\ emission lines really absent in AGNs?  If so, why?

It is possible that narrow \feii\ lines are actually prevalent in the optical
spectra of AGNs, yet have just been neglected or mistaken as a part of the
co-existing broad-line emission or stellar features.  Because of the complex
atomic structure of the Fe$^+$ ion, there are so many transitions that
individual \feii\ lines, even if narrow, are highly blended with each other,
effectively mimicking much broader features.  The situation is further
exacerbated in type~1 AGNs by the ubiquitous presence of broad \feii\
multiplets as well as other emission lines.  Without knowing a complete
identification list of the optical transitions and their relative strengths,
it is hard to discern them.  Fortunately, V\'eron-Cetty et al. (2004) have
recently identified and measured all the narrow \feii\ lines present in the
optical spectrum of I~Zw~1, a well-known narrow-line Seyfert 1 galaxy.
Their analysis made it possible for us to explore systematically narrow
optical \feii\ emission lines in AGNs.

In this Letter, we report the discovery of the prevalence of narrow \feii\
emission lines, both permitted and forbidden, in the optical spectra of
type~1 AGNs and their nondetection in type~2 AGNs.  We adopt a cosmology
with $H_{\rm 0}$ = 70 km\,s$^{-1}$\,Mpc$^{-1}$, $\Omega_{\rm m}$ = 0.3, and
$\Omega_{\Lambda}$ = 0.7.

\section{Sample and Data Analysis}

\subsection{The Type 1 AGN Sample}

The type~1 AGN sample consists of the 4178 Seyfert 1 galaxies and quasars from
Dong et al. (2010), selected from the spectral data set of the Sloan Digital
Sky Survey Fourth Data Release (SDSS DR4; Adelman-McCarthy et al. 2006).
Sample definition and data analysis methods are described in detail in that work.
Here we provide a brief description and some additional treatments
for verifying the existence of narrow-line \feii\ emission.
Briefly, we select broad-line AGNs
with prominent \feii\ emission in the SDSS bandpass and
with continuum and emission lines suffering minimally from contamination
by host galaxy starlight.  The criteria are as follows: (a) $z \leq 0.8$,
to ensure that
\feii\ emission in the restframe wavelength region 4434--4684 \AA\
(hereinafter \feii\,$\lambda4570$),
as well as H$\beta$ and \oiii\,$\lambda\lambda$4959, 5007, are in the bandpass;
(b) a median signal-to-noise ratio (S/N) $\geq 10$ per pixel
in the optical \feii\ region (4400--5400 \AA); and (c) absence of detectable
Ca\,K (3934 \AA), Ca\,H + H$\epsilon$ (3970 \AA), and H$\delta$ (4102 \AA)
stellar absorption features at $> 2\,\sigma$ significance.

At optical wavelengths numerous \feii\ multiplets and other broad emission
lines are heavily blended to form a pseudocontinuum.  Following Dong
et al. (2008), we fit simultaneously the AGN featureless continuum,
represented by a power law, the \feii\ multiplets, and other emission lines
in the range of 4200--5600 \AA\ using a code based on the MPFIT package
(Markwardt 2009).  The \feii\ emission is modeled with two separate sets of
templates in \emph{analytical} forms,%
\footnote{The implementation of the template functions in Interactive Data Language (IDL)
is available at
http://staff.ustc.edu.cn/\~{ }xbdong/Data\_Release/FeII/Template/ \,.}
one for the broad-line system and the other for the narrow-line system,
constructed from measurements of I~Zw~1 by V\'eron-Cetty et al. (2004), as
listed in their Tables~A1 and A2.  Within each system, the respective set of
\feii\ lines are assumed to have no relative velocity shifts and the same
relative strengths as those in I~Zw~1.  In the optical spectrum of I~Zw~1, the
broad \feii\ lines are blueshifted by 150 \kms, whereas the narrow \feii\
lines are located at the systematic redshift of the host galaxy; this
velocity separation made possible the identification of the narrow component
from the broad one (P. V\'eron 2009, private communication).
The broad \hb\ line is fitted with as many Gaussians as
statistically justified.  All narrow emission lines are fitted with a single
Gaussian, except for the \oiii\,$\lambda \lambda$4959, 5007 doublet, each of
which is modeled with two Gaussians, one accounting for the line core and the
other for a possible blue wing, as seen in many objects.  The redshift and
width of narrow-line \feii\ are set as free parameters.

In order to verify that the narrow-line \feii\ are not residuals from poor
broad \feii\ subtraction due to mismatch of the broad-line \feii\ model,
we use three schemes to model the profile of individual broad \feii\ lines:
(A) the best-fit broad \hb\ profile,
(B) the best-fit total (narrow + broad) \hb\ profile, and
(C) a single Lorentzian with width set as a free parameter.
Scheme A is a natural choice that has been used in our previous studies
(e.g., Dong et al. 2008, 2010).  Scheme B is rather extreme and formally
unreasonable because \hb\ definitely has a component from the canonical NLR;
we use it as a stringent test of the reality of the narrow \feii\ lines.
Scheme C is inspired by the fact that the broad \feii\ lines in I~Zw~1 are
best described by a Lorentzian profile (V\'eron-Cetty et al. 2004).
The redshift of broad \feii\ is set to be a free parameter.
We note that the width of broad \feii\ lines may be different from that of broad \hb\
(see, e.g., Hu et al. 2008),
yet we find that the flux of both broad and narrow \feii\ are quite insensitive to the exact
profile assumed for the broad component (Dong et al. 2010).
As noted by Vestergaard \& Peterson (2005) and Landt et al. (2008), the broad \feii\
multiplets are so highly blended that the overall profile 
mainly depends on their relative strengths.
As a strict demonstration
that the detection of narrow \feii\ is robust, we show (\S3.1) that we
can recover these features in the residual spectrum even after excluding
the narrow-line \feii\ template from our model altogether.

\subsection{The Type 2 AGN Sample}

The type~2 AGN sample comprises the $\sim 27,000$ Seyfert 2 galaxies in
SDSS DR4 selected according to the criteria of Kauffmann et al. (2003), having
\hb, \oiii\,$\lambda 5007$, \ha, and \nii\,$\lambda6583$ detected at
$> 5\,\sigma$ significance.  The mean and standard deviation of their
redshifts are 0.10 and 0.05, respectively.  We subtracted the starlight
continuum to obtain a clean emission-line spectrum using the method of
Lu et al. (2006), using stellar templates broadened and shifted to
match the stellar velocity dispersion of the galaxy.
The stellar absorption lines must be subtracted well to ensure reliable
measurement of weak emission lines.  In the present study, we add the
narrow \feii\ template described above into the model, in order to detect any
possible narrow \feii\ emission.  Next, we fitted emission lines with
Gaussians using the code described in detail in Dong et al. (2005); errors
are given by MPFIT.  There are 2671 objects in the galaxy catalog having a
broad \ha\ component with S/N $>5$; we excluded these from the type~2 AGN
sample (cf. Zhang et al. 2008).

\section{Results}

\subsection{Narrow \feii\ Emission in Type 1 AGNs}

As a first step toward ascertaining whether narrow \feii\ lines can be
discerned without the aid of a narrow \feii\ template, we model the
pseudocontinuum with a power-law AGN continuum plus the broad \feii\
template only.  For this purpose, we try all three schemes of modeling the
profile of broad \feii\ lines described in \S2.1.  The fitting results are
very encouraging: narrow \feii\ lines are present in many residual spectra
regardless of the broad \feii\ model adopted.  Figure 1 illustrates the
three schemes applied to SDSS\,J113541.19+002235.4.  Plotted  as vertical
lines are the locations of the narrow \feii\ lines, both permitted (dotted) and
forbidden (dashed), as well as some narrow transitions of \crii, \niii, and
\tiii\ (cyan), that were identified by V\'eron-Cetty et al. (2004)
and appear unambiguous in all the residual spectra.
Those lines are also labeled in Figure 1.
While there are
small differences among the three fitting schemes, overall the pattern of
narrow features appears robust.  Moreover, the relative strengths of the
narrow features seem, to first order, reproducible from  object to object.
This is shown in Figure 2 (top), where we plot the residual spectra
for four sources fitted using Scheme A (broad \feii\ modeled using the broad \hb\
profile).

Next, we fit the entire type~1 sample with the narrow \feii\ template
included into the pseudocontinuum model, as described in \S2.1.
The broad \feii\ line profile is again modeled with the above three schemes.
The addition of the narrow-line component is justified statistically by the
$F-$test,%
\footnote{We have found through experimentation that the $F-$test works well
in such spectral fitting, although, theoretically, it holds only for linear
models (see below and Dong et al. 2008).}
with over half of the spectra having a chance probability $\pnull < 0.01$,
regardless of the model adopted for the broad \feii\ profile (even the
extreme case of Scheme B).  Because the fluxes of both narrow  and broad
\feii\ are relatively insensitive to the choice of broad \feii\ template,
we adopt Scheme A (broad \feii\ profile = broad \hb\ profile) for the rest
of the study.

We detect narrow \feii\ emission  at $> 3\,\sigma$ significance in 2515
(60\%) of the 4178 objects, and at $> 5\,\sigma$ significance in 1872 (45\%),
using errors on the narrow-line \feii\ flux as given by MPFIT.  This is
consistent with the above $F-$test results.  The restframe equivalent widths
of the narrow \feii\ features integrated over the region 4434--4684 \AA\
(\feii$^{\rm N}$\,$\lambda4570$) range from undetectable (\lax 0.1 \AA) to 25 \AA.
The intensity ratios of the narrow to broad \feii\,$\lambda4570$
feature, among the 2502 objects in which both components are detected at
$>3\,\sigma$ significance, vary by 2 orders of magnitude, ranging from
$\sim$0.005 to 0.5, with a mean of 0.07 and a standard deviation of 0.3 dex
(computed in log-space).  The equivalent widths of the two components do not
correlate very tightly (Spearman correlation coefficient $\rs = 0.45$; see Figure 3),
suggesting that the narrow component is not
an artifact of measurement uncertainty associated with deblending of the broad
component.  Remarkably, however, as described in the companion paper
by Dong et al.  (2010), the relative strengths of narrow \feii\
with respect to the continuum and all other
prominent emission lines in the near-ultraviolet and optical region (e.g.,
\mgii\,$\lambda 2800$, broad \feii, broad \hb, and \oiii\,$\lambda 5007$)
correlate most strongly with the Eddington ratio%
\footnote{Eddington ratio, $\ell \equiv \lratio$, is the ratio between
the bolometric and Eddington luminosities.
The Eddington luminosity (\ledd), by definition, is the
luminosity at which the gravity of the central source acting on an
electron--proton pair (i.e., fully ionized gas) is balanced by the radiation
pressure due to electron Thomson scattering.
}
rather than with other physical parameters of the AGNs.
A physical model unifying these and other correlations concerning Eddington ratio
were proposed by Dong et al. (2009a, 2009b, 2010).

It is difficult to obtain accurate measurements of the kinematics of the narrow
\feii\ features.  The lines are weak, most are severely blended,
and both the spectral resolution and S/N of the SDSS data are limited.
Notwithstanding these limitations, we make a preliminary statistical investigation
for the 1032 objects with narrow \feii\ emission detected at $> 10\,\sigma$ significance.
The best-fitting narrow \feii\ model gives line widths
(full width at half-maximum, corrected for instrumental broadening) mostly in the range
of 200--800 \kms, with a mean of 560 \kms\
and a standard deviation of 150 \kms; these values are comparable
to those of \oiii\,$\lambda$5007 within $\pm 50\%\,\,(1\,\sigma)$.
The velocity shifts of the narrow \feii\ system with respect to narrow \hb\ range mostly
from $-200$ \kms\ (blueshifted) to 300 \kms\ (redshifted), with a mean of 50 \kms\ and
a standard deviation of 100 \kms.
With respect to the broad \feii\ system, the mean velocity shift of narrow \feii\
is 210 \kms, with a standard deviation of 280 \kms.
We cannot discern any obvious velocity differences between the permitted and forbidden
narrow transitions. Without a robust estimation of the measurement uncertainties on these
kinematic parameters at present, we refrain from a full discussion of the kinematics
in this Letter.

\subsection{No Narrow \feii\ Emission in Type 2 AGNs}

In strong contrast to the case of type~1 AGNs, no \feii\ emission is detected
at $\geq 3\,\sigma$ significance {\em in any object in the type~2 sample!}
Considering that the detection of weak emission lines may be heavily
influenced by uncertainties in starlight subtraction, we divide the type~2
sample into three subsamples according to \oiii\,$\lambda5007$ luminosity,
and build a composite spectrum for each.  The composites are arithmetic mean
spectra, constructed following Chen et al. (2010; see their \S3.1). The luminosity ranges
of the three subsamples are $L_{\rm [O III]} < 10^{40.5}$, $10^{40.5}- 10^{41.5}$,
and $> 10^{41.5}$ \lum.  To avoid possible confusion from the inclusion of
star-forming galaxies, we further require the objects to be located above the
maximum starburst line in the Baldwin-Phillips-Terlevich diagram (Baldwin
et~al. 1981) of Kewley et al. (2001), and to have \oiii\,$\lambda5007 / \hb
> 3$ to avoid low-ionization sources.  To obtain a comparison sample of
type~2 sources with AGN luminosities more compatible with that of the type~1
sample (median $L_{\rm [O III]} = 1.3\times 10^{42}$ \lum), we also build a
composite spectrum for the SDSS type~2 quasars presented in Reyes et al.
(2008), which have $8 \times 10^{41} < L_{\rm [O III]} <  4 \times 10^{43} $ \lum.
As shown in Figure 2 (bottom), it is clear that no narrow \feii\ features
whatsoever are seen in any of the composite type~2 spectra.
To quantify this striking result, we use the type~2 quasar composite, after starlight
subtraction, to place upper limits on two relatively isolated features,
\feii\,$\lambda4925$ (integrated in the vacuum wavelength range 4918--4938 \AA,
which is dominated by \feii\,42 $\lambda4923$ and \feii]\,$\lambda4928$)
and \feii\,49 $\lambda5234$.
Relative to \oiii\,$\lambda$5007, the limits are $< 0.006$ and $< 0.005$, respectively.
By comparison, the mean values of these features in the type~1
sample are $0.05$ and $0.02$.

\section{Discussion and Summary}

The most surprising finding of this Letter is that narrow \feii\ emission is
prevalent in type~1 AGNs, yet not present at all in type~2 AGNs.  This is
a statistically robust result, based on analysis of a large sample of
individual spectra as well as very high-S/N stacked spectra.  A possible
explanation for this striking contrast, appealing to the canonical geometric
unification scheme for type~1 and type~2 AGNs (e.g., Antonucci 1993), is that
\emph{narrow \feii\ emission arises from gas in the innermost regions of the
NLR located interior to the obscuring torus}.  The inner edge of the torus
has been estimated to be on scale of parsecs, roughly the dust sublimation
radius (Suganuma et al. 2006); its radial extent is likely to be several tens
of parsecs (e.g., Jaffe et al. 2004; see Granato \& Danese 1994 for a model).
In this picture the narrow \feii-emitting region is visible along our
line-of-sight in type~1 objects but obscured by the (extent of) the dusty
torus in type~2 counterparts.  As a low-ionization specie, \feii\ may
preferentially avoid the ionization cone and be largely confined to a
disk-like geometry along the plane of the torus, as depicted in Figure 6 of
Gaskell (2009).

Yet, the complete absence of optical narrow \feii\ emission in type~2 sources
appears perplexing in view of the fact that narrow [\feii] emission in the
NIR has been seen in AGNs of both types (see \S1). Can line-of-sight dust
obscuration be so effective in suppressing or hiding the optical lines,
in all cases?  Given the inherent patchiness of the interstellar medium, this
seems incredible.  However, in the absence of a quantitative prediction for
the intrinsic relative strengths of the optical and NIR \feii\ transitions,
which, to our knowledge, has not been published, it is difficult to evaluate
this hypothesis.  Comparison between the NIR and optical lines may be
further complicated by the fact that the transitions in these two wavelength
ranges arise from different energy levels ($< 1.5$ eV for the NIR; $\sim 3$ eV
for optical forbidden; $\sim 5$ eV for optical permitted), and hence likely
trace physically different regions.  The NIR lines probably tracer cooler,
dustier material compared to the regions emitting the optical lines.

There are several lines of fruitful work for the future.  From an
observational standpoint, it would be useful to obtain optical spectra of
higher S/N and higher spectral resolution to better constrain the kinematics
of the \feii-emitting gas and its relation to regions emitting
higher-ionization species.  To better understand the puzzling mismatch
between the optical and NIR lines, it would be instructive to obtain
NIR spectra of the sources for which narrow \feii\ has been detected in
the optical, and vice versa.  And finally, theoretical calculations
are needed to predict the intrinsic spectrum of narrow \feii\ across a
wide wavelength range.

\acknowledgments
We thank Kirk Korista, Martin Gaskell, and Weimin Yuan for helpful comments.
This work is supported by Chinese NSF grants NSF-10703006 and NSF-10728307,
and a National 973 Project of China (2007CB815403).
The research of L.C.H. is supported by the Carnegie Institution for Science.
The visit of X.D. in Carnegie Observatories is mainly supported by OATF, USTC.
Funding for the SDSS and SDSS-II has been provided by the Alfred P. Sloan
Foundation, the Participating Institutions, the National Science Foundation,
the U.S. Department of Energy, the National Aeronautics and Space
Administration, the Japanese Monbukagakusho, the Max Planck Society,
and the Higher Education Funding Council for England.
The SDSS Web Site is http://www.sdss.org/.



\figurenum{1}
\begin{figure}[tbp]
\epsscale{0.75}
\plotone{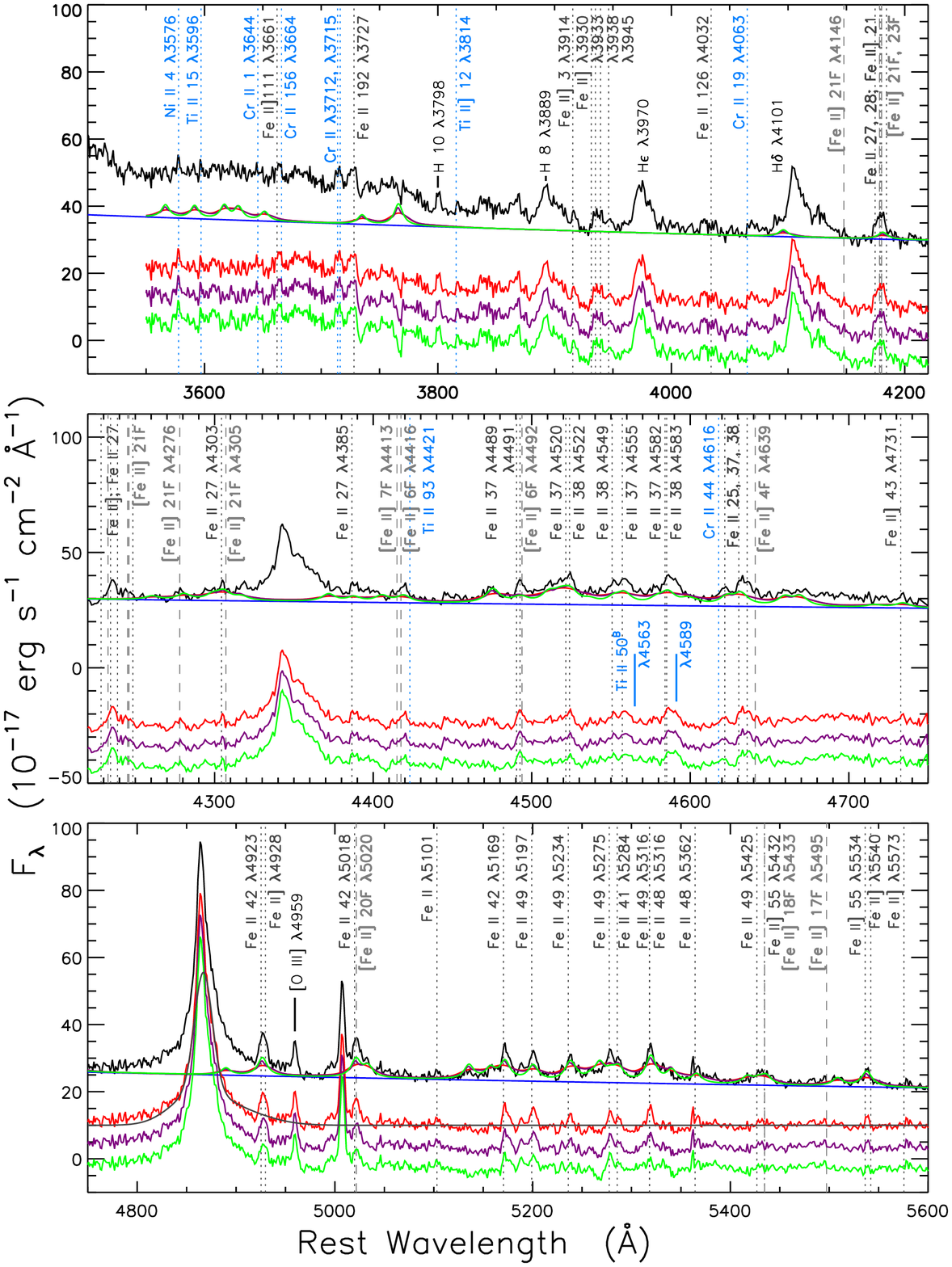}
\caption{
Fitting results of the optical spectrum of SDSS\,J113541.19+002235.4, using
three different models for the profile of broad \feii\ emission lines,
without taking into account narrow \feii\ lines.  We show the SDSS spectrum
(black), the AGN power-law continuum (blue), the pseudocontinuum (AGN
continuum plus broad-line \feii\ emission) with broad \feii\ lines modeled
with the profile of broad \hb\ (red), with the whole \hb\ profile (purple),
and with a single Lorentzian of variable width (green).  In the bottom of
every panel, the pseudocontinuum-subtracted residuals for the three
respective models are displayed (with arbitrary offsets for clarify).
Vertical lines and labels denote the permitted (dark gray, dotted) and forbidden
(gray, dashed) narrow \feii\ lines,
as well as some narrow \crii, \niii\ and \tiii\ lines (cyan),
that appear unambiguous in the present spectrum.
We also label several non-iron emission lines (black) and two broad \tiii\
lines (navy).  In the bottom panel we show the best-fit model for broad
\hb\ (gray, solid).
}
\end{figure}

\figurenum{2}
\begin{figure}[tbp]
\epsscale{0.95}
\plotone{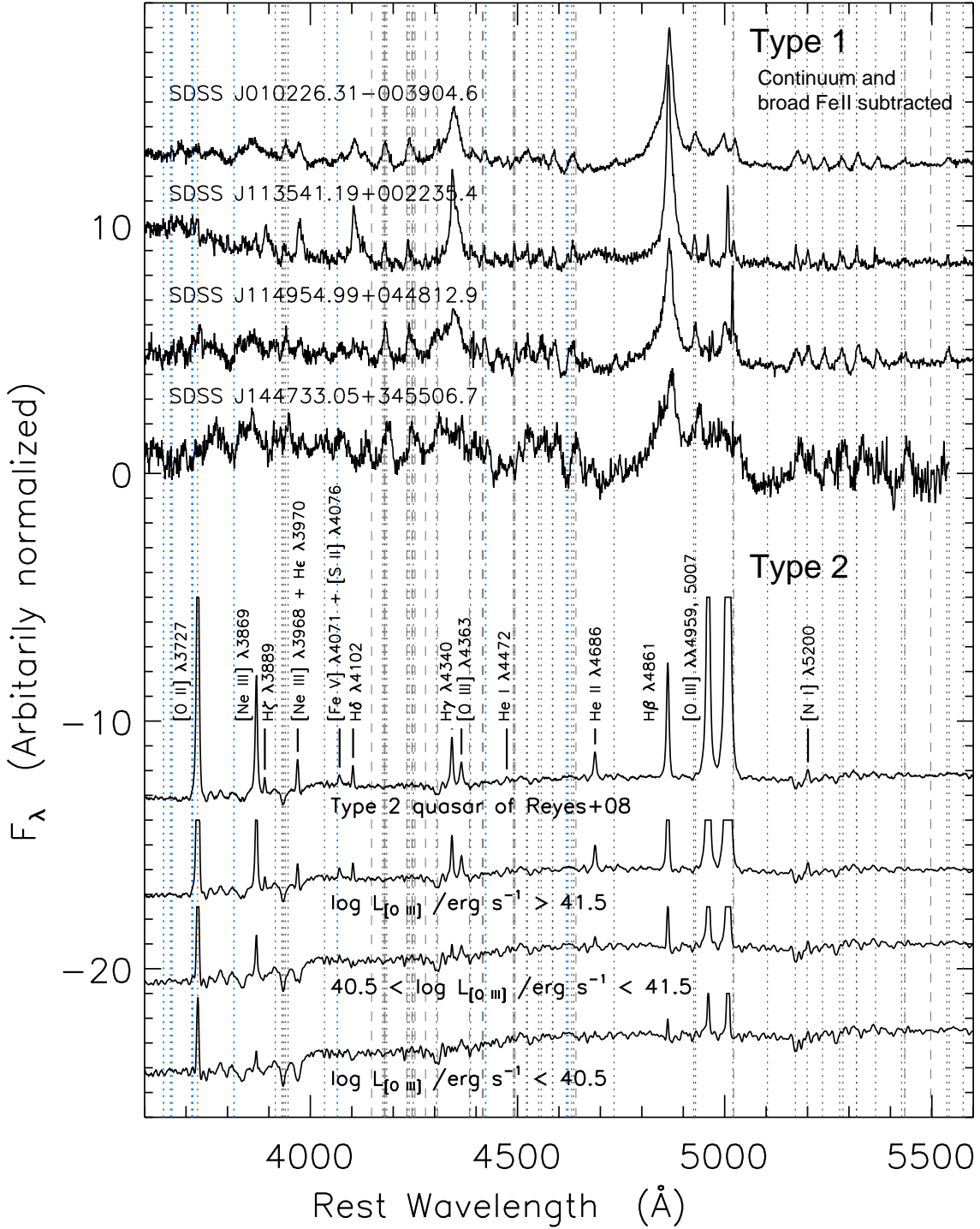}
\caption{
({\it Top}) Residual spectra of four type~1 AGNs with strong narrow \feii\
lines, after the subtraction of the AGN continuum and the best-fit broad-line
\feii\ emission modeled with the profile of broad \hb.  The vertical lines
are the same as in Figure 1.  ({\it Bottom}) Four
composite spectra of type~2 AGNs, ordered by \oiii\ luminosity.  The narrow
emission lines from non-iron elements are labeled according to the
identifications by Vanden Berk et al. (2001) from their composite quasar
spectrum.  Strong emission lines are clipped for clarity.
}
\end{figure}

\figurenum{3}
\begin{figure}[tbp]
\plotone{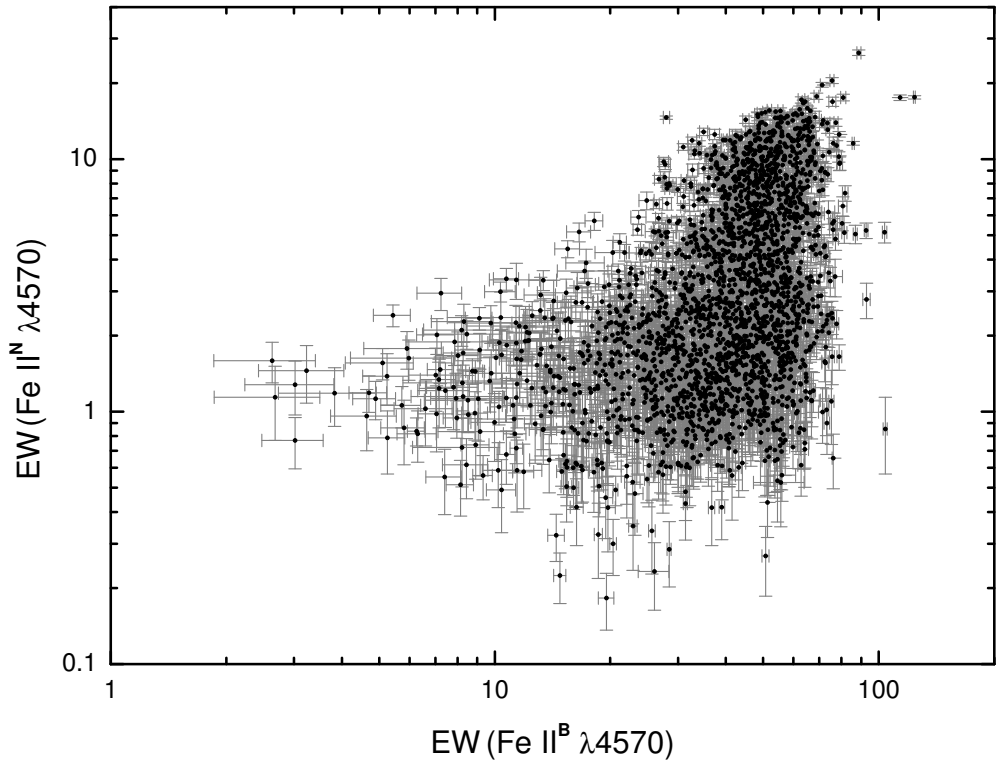}
\caption{
Distribution of equivalent widths of narrow and broad \feii\,$\lambda4570$
emission (black dots), along with $1\,\sigma$ error bars (gray), for the 2502
objects with both components detected at $>3\,\sigma$ significance.
}
\end{figure}

\end{document}